\documentclass[11pt]{article}

\usepackage{amsmath}
\usepackage{graphicx}
\usepackage{indentfirst}
\usepackage{amssymb}
\usepackage{cite}
\usepackage{color}
\usepackage{subfigure}

\begin{document}

\title{\bf{Weak Cosmic Censorship with Pressure and Volume\\ in Charged Anti-de Sitter Black Hole\\ under Charged Scalar Field}}

\date{}
\maketitle

\begin{center}
\author{Bogeun Gwak}$^a$\footnote{rasenis@sejong.ac.kr}\\

\vskip 0.25in
$^{a}$\it{Department of Physics and Astronomy, Sejong University, Seoul 05006, Republic of Korea}\\
\end{center}
\vskip 0.6in

{\abstract
{We investigate the weak cosmic censorship conjecture in charged anti-de Sitter black holes with thermodynamic pressure and volume by the scattering of the charged scalar field in four and higher dimensions. We assume that the internal energy and electric charge of the black hole changes infinitesimally according to the energy and charge fluxes of the scalar field. Then, found to be an isobaric process, the changes in the black hole can be well reproduced to the first law of thermodynamics, although we find that the second law of thermodynamics is violated in extremal and near-extremal black holes. Nevertheless, the weak cosmic censorship conjecture remains valid, because the extremality of a black hole is invariant despite changes in internal energy and electric charge.}}

\thispagestyle{empty}
\newpage
\setcounter{page}{1}

\section{Introduction}\label{sec1}

Black holes are massive objects in the universe, and they exhibit various features of gravity. A significant feature of a black hole is its event horizon, through which no matter is able to escape. This is also applicable to light, such that an outside observer cannot detect any radiation from the inside of a black hole. From the perspective of quantum physics, there is an energy radiated from the black hole, such that, according to Hawking radiation, black holes can be treated as a thermodynamic system that remains in Hawking temperature\cite{Hawking:1974sw,Hawking:1976de}. The mass of a black hole can be divided into two types: irreducible mass, and reducible energy\cite{Christodoulou:1970wf,Bardeen:1970zz,Christodoulou:1972kt}. The irreducible mass increases in any irreversible process. However, the mass of a black hole can decrease, as with the Penrose process\cite{Penrose:1971uk}. Here, reduced mass is the reducible energy among the energy of a black hole. This reducible energy includes electric and rotational energies, and it can be reduced by external fields or particles. In thermodynamics, the irreducible property of entropy is similar to that of irreducible mass, and the Bekenstein--Hawking entropy of a black hole is given to be proportional to the square of the irreducible mass\cite{Bekenstein:1973ur,Bekenstein:1974ax}. According to these definitions of the temperature and entropy of a black hole, the laws of thermodynamics are defined.

Recently, the thermodynamic pressure and volume of the black hole can be introduced into the laws of thermodynamics. Here, pressure is defined by the cosmological constant as a dynamic variable. In fact, the dynamic cosmological constant is a concept that has been assumed for some time \cite{Teitelboim:1985dp,Brown:1988kg}. Since then, the cosmological constant has been introduced to reflect pressure in the spacetime of a black hole\cite{Caldarelli:1999xj, Padmanabhan:2002sha}. According to the laws of thermodynamics, furthermore, a thermodynamic conjugate must be introduced to the pressure. This conjugate is the thermodynamic volume of the black hole\cite{Dolan:2010ha,Cvetic:2010jb}. The combination extends the first law of thermodynamics with the $PV$ term\cite{Dolan:2011xt}. Here, the mass of a black hole does not correspond to its internal energy, but rather to the enthalpy\cite{Kastor:2009wy}. Since the physical implication of the mass changes from internal energy to enthalpy, thermodynamic phenomena differ depending on whether the $PV$ term is included. In particular, various studies have focused on thermodynamic applications with the $PV$ term\cite{Kubiznak:2012wp,Johnson:2014yja,Caceres:2015vsa,Karch:2015rpa,Kubiznak:2016qmn,Hennigar:2017apu,Wei:2017vqs,Gregory:2017sor}.

The asymptotic geometry of a black hole with the negative cosmological constant is anti-de Sitter (AdS) spacetime. For AdS spacetime, the gravity solution in a $D$-dimensional bulk corresponds to conformal field theory (CFT) on the $D-1$-dimensional boundary\cite{Maldacena:1997re,Gubser:1998bc,Witten:1998qj,Aharony:1999ti}. This is the well-known AdS/CFT correspondence. According to this correspondence, physics in the AdS bulk has an implication for CFT on its boundary. Particularly, AdS black holes play a significant role when introducing thermodynamic properties to the corresponding CFT\cite{Witten:1998zw}. For example, an AdS black hole corresponds to finite-temperature CFT, but the AdS spacetime corresponds to zero temperature. According to AdS/CFT correspondence, AdS solutions are connected to various physical situations. As a representative application, an extended correspondence is found to associate with condensed matter theory (CMT), referred to as AdS/CMT correspondence\cite{Hartnoll:2008vx,Hartnoll:2008kx}. Here, the correspondence to the charged AdS black hole is found to be holographic superconductors\cite{Ren:2010ha,Jensen:2010em,Andrade:2011sx,Haehl:2013hoa,Chang:2014jna}.

Solutions to black holes have a singularity at their inside. Indeed, the existence of this singularity is inevitable according to Einstein's theory of gravity \cite{Hawking:1969sw}. Since a naked singularity without a horizon causes problems in terms of causality, the cosmic censorship conjecture states that the singularity should be hidden to an observer in the spacetime of a black hole\cite{Penrose:1964wq,Penrose:1969pc}. There are two forms of this conjecture, depending on the kind of observer. The weak cosmic censorship (WCC) conjecture, our main focus, suggests that the singularity is invisible to an asymptotic observer owing to the horizon. Hence, the horizon should be stable. The other is the strong cosmic censorship (SCC) conjecture, in which the singularity is invisible to all observers. The first test of the WCC conjecture involved adding a particle to the Kerr black hole\cite{Wald:1974ge}. Regarding extremal Kerr black holes, this test demonstrated the validity of the WCC conjecture. Notably, there is no generalized proof for a test of the WCC conjecture, so each investigation can come to a different conclusion depending on its assumptions. For instance, the horizon of a near-extremal Kerr black hole can be unstable with the addition of a particle\cite{Jacobson:2009kt}. This invalidity differs from the validity supposed in an extremal black hole. Subsequently, if self-force or back-reaction is introduced, the instability is resolved, and the WCC conjecture becomes valid for near-extremal black holes\cite{Barausse:2010ka,Barausse:2011vx,Colleoni:2015ena,Colleoni:2015afa,Sorce:2017dst}. Similar studies have been conducted in Reissner--Nordstr\"{o}m black holes with a particle\cite{Hubeny:1998ga,Isoyama:2011ea}. Indeed, the WCC conjecture is continuously tested in various black holes\cite{BouhmadiLopez:2010vc,Gwak:2011rp,Rocha:2011wp,Zhang:2013tba,Cardoso:2015xtj,Siahaan:2015ljs,Horowitz:2016ezu,Gwak:2016gwj,Revelar:2017sem,Song:2017mdx,Gwak:2018tmy,Shaymatov:2018fmp,Gim:2018axz,Chen:2018yah,Duztas:2018ebr}. We can expect that the existence of the $PV$ term affects the validity of the WCC conjecture when adding a particle. Fortunately, when testing a charged AdS black hole, the validity of the WCC conjecture was demonstrated with the $PV$ term\cite{Gwak:2017kkt}.

The WCC conjecture can also be investigated under the scattering of external fields, rather than through the addition of a particle. Here, the responses in the black hole depend on the scattered field contents, such as its modes and spin number, which are not considered when adding a particle. In consideration of scalar and Maxwell fields, various tests for the WCC conjecture were conducted \cite{Hod:2008zza,Semiz:2005gs,Toth:2011ab,Natario:2016bay,Duztas:2017lxk}. For a Kerr--(anti-)de Sitter black hole, the WCC conjecture was proven valid in both extremal and near-extremal cases, because there is a limit to the transfer of energy from the scalar field to the black hole during a given time interval\cite{Gwak:2018akg}. Actually, the scattering of an external field has distinct features compared to cases of a particle. The superradiance is a process when extracting the energy of a black hole from the radiation as it scatters with an external field given under specific conditions\cite{Zeldovich:1971aa,Zeldovich:1972ab}. Since the radiation can be amplified according to the asymptotic geometry, it is associated with the instability of a black hole. In cases of AdS black holes, the instability depends on their size under superradiance. For example, small Kerr-AdS black holes are unstable, but large ones are stable\cite{Hawking:1999dp,Cardoso:2004hs,Cardoso:2006wa,Uchikata:2009zz,Cardoso:2013pza,Delice:2015zga}. Superradiance has been studied recently in various AdS black holes\cite{Aliev:2008yk,Aliev:2015wla,Dias:2016pma,Ferreira:2017tnc,Sullivan:2017agx,Rahmani:2018hgp}. Further, the scattering of external fields plays an important role when testing the SCC conjecture, and the decay rate of its quasinormal modes is significant.

In this paper, we investigate the WCC conjecture in $D$-dimensional charged AdS black holes in consideration of thermodynamic pressure and volume by the scattering of the charged scalar field. To our knowledge, this is the first study of the WCC conjecture with the $PV$ term {\it by the charged scalar field}. Since changes in the black hole differ considerably  given the addition of the $PV$ term, we expect that the $PV$ term will be significant to the laws of thermodynamics and the WCC conjecture. We begin with the Lagrangian for a nonminimally coupled massive scalar field with an electric charge. Then, we obtain the energy and charge carried into the black hole in terms of the energy and charge fluxes of the scalar field at the outer horizon. According to the fluxes, the state in the black hole can be estimated from the initial state after an infinitesimal time interval. Under this process, the changes in the black hole come together to produce the first law of thermodynamics with the $PV$ term. However, the second law of thermodynamics is {\it violated}, because Bekenstein--Hawking entropy is reduced when the initial state is a near-extremal black hole (including an extremal black hole). This violation is only observed in the case with the $PV$ term. Then, the WCC conjecture is investigated in a near-extremal black hole. Under the scattering, the mass and charge of the black hole change as much as those transferred by the fluxes of the scalar field, but the initial extremality is invariant. This implies that the extremal black hole is still {\it extremal}, and that a near-extremal black hole remains {\it near-extremal}. Even if the second order of the variation is considered, the black hole cannot be overcharged beyond the extremal condition. This is because we consider the scattering of the charged scalar field under which the carried energy and charge have a limit during the infinitesimal time interval. Therefore, by scattering the charged scalar field, we prove the {\it validity} of the WCC conjecture. Further, we apply our analysis to the saturation of the black hole with the potential of the scalar field. Based on the changes in the black hole, our analysis shows that the saturation of the black hole takes the very long time. 

This paper is organized as follows: Section\,\ref{sec2} reviews the charged AdS black hole and thermodynamics with the thermodynamic pressure and volume; Section\,\ref{sec3} obtains the charged scalar field equations at the horizon of the charged AdS black hole; Section\,\ref{sec4} tests the laws of thermodynamics with the $PV$ term in the charged AdS black hole; Section\,\ref{sec5} validates the WCC conjecture in extremal and near-extremal AdS black holes with the $PV$ term; Section\,\ref{sec6} finds the change in the electric potential of the charged AdS black hole under the scattering of the charged scalar field; and Section\,\ref{sec7} summarizes our results.

\section{Thermodynamics with Pressure and Volume in AdS Black Holes}\label{sec2}

We consider an electrically charged AdS black hole in higher dimensions (including four dimensions). We derive a solution to Einstein--Maxwell gravity theory with a negative cosmological constant in $D$-dimensional spacetime. The action is given as
\begin{align}\label{eq:action01}
S=-\frac{1}{16\pi}\int d^D x \sqrt{-g} \left(R-F_{\mu\nu}F^{\mu\nu}-2\Lambda\right),
\end{align}
where the curvature is denoted by $R$. The field strength $F_{\mu\nu}$ and cosmological constant $\Lambda$ are defined in terms of the gauge field $A_\mu$ and AdS radius $\ell$ as
\begin{align}\label{eq:action02a}
F_{\mu\nu}=\partial_\mu A_\nu - \partial_\nu A_{\mu}, \quad  A_\mu = -\delta_\mu^0 \frac{Q}{r^{D-3}},\quad \Lambda=-\frac{(D-1)(D-2)}{2\ell^2}.
\end{align} 
Then, the solution to the charged AdS black hole is obtained from field equations of the action in Eq.\,(\ref{eq:action01}). The metric is 
\begin{align}\label{eq:BHmetric01a}
ds^2 = -\frac{\Delta}{r^2}dt^2 + \frac{r^2}{\Delta}dr^2 +r^2 d\Omega_{D-2},\quad \Delta(r)= r^2-\frac{2M}{r^{D-5}}+\frac{Q^2}{r^{2D-8}}+\frac{r^4}{\ell^2},
\end{align}
where $M$ and $Q$ are mass and electric charge parameters, respectively. The metric and surface area of $(D-2)$-sphere $\Omega_{D-2}$ are denoted by 
\begin{align}\label{eq:BHmetric01b}
d\Omega_{D-2}=\sum^{D-2}_{i=1}\left( \prod_{j=1}^{i} \sin^2\theta_{j-1}\right)d\theta_i^2,\quad \theta_0\equiv\frac{\pi}{2},\quad \theta_{D-2}\equiv\phi,\quad \Omega_{D-2}=\frac{2\pi^\frac{D-1}{2}}{\Gamma(\frac{D-1}{2})}.
\end{align}
There are two horizons satisfying $\Delta(r)=0$: the inner horizon, and the outer horizon. Here, we focus on the WCC conjecture and thermodynamics, so the outer horizon is mainly considered and denoted by $r_\text{h}$. Note that the metric in Eq.\,(\ref{eq:BHmetric01a}) becomes a Reissner--Nordtr\"{o}m AdS black hole in four dimensions. The mass and electric charge of the black hole depend on the dimensionality of the spacetime \cite{Tian:2010gn}.
\begin{align}
M_\text{B}=\frac{(D-2)\Omega_{D-2}}{8\pi}M,\quad Q_\text{B}=\frac{(D-2)\Omega_{D-2}}{8\pi}Q.
\end{align}
Here, we consider the extended thermodynamics given in a $D$-dimensional charged AdS black hole. The extended thermodynamics include the pressure and volume, where the cosmological constant is defined for the thermodynamic pressure $P$. The cosmological constant is considered a fixed value in the action of Eq.\,(\ref{eq:action01}), and its mathematically incompleteness is clear. In this extension, the cosmological constant is an effective value originating from the expectation value of gravity theory\cite{Cvetic:2010jb}. This implies that the variation of the cosmological constant is effective shorthand for denoting the decay into a different vacuum expectation value. The thermodynamic volume $V_\text{B}$ is defined as the conjugate variable of the pressure. According to the $PV$ term, the extended thermodynamics is well constructed\cite{Dolan:2010ha,Dolan:2011xt}. The definitions for thermodynamic pressure and volume are \cite{Dolan:2013ft}
\begin{align}
P=-\frac{\Lambda}{8\pi}=\frac{(D-1)(D-2)}{16\pi \ell^2},\quad V_\text{B}=\frac{\Omega_{D-2}}{D-1}r_\text{h}^{D-1}.
\end{align}
The Hawking temperature of the black hole is given as
\begin{align}\label{eq:temperature03a}
T_\text{h}=\frac{1}{4\pi}\left(\frac{d\Delta_\text{h}}{r_\text{h}^2}\right)=\frac{1}{4\pi r_\text{h} \ell^2 }\left((D-1)r_\text{h}^2+(D-3)\ell^2-\frac{(D-3)Q^2\ell^2}{r_\text{h}^{2D-6}}\right),\quad d\Delta_\text{h}=\frac{\partial \Delta}{\partial r}\Big|_{r=r_\text{h}}.
\end{align}
Further, Bekenstein--Hawking entropy and electric potential are given at the outer horizon as
\begin{align}\label{eq:temperature03a}
S_\text{h}=\frac{\Omega_{D-2}r_\text{h}^{D-2}}{4},\quad \Phi_\text{h}=\frac{Q}{r_\text{h}^{D-3}}.
\end{align}
When we consider the thermodynamic pressure and volume, the mass of the black hole is defined as its enthalpy \cite{Kastor:2009wy,Cvetic:2010jb}.
\begin{align}\label{eq:enthalpy01}
M_\text{B}=U_\text{B}+PV_\text{B},
\end{align}
where the internal energy $U_\text{B}$ plays an important role in our analysis of the WCC conjecture. Then, the change in the mass of the black hole is given by the first law of thermodynamics with the terms for pressure and volume \cite{Dolan:2012jh,Dolan:2013ft}.
\begin{align}
dM_\text{B}=T_\text{h}dS_\text{h}+\Phi_\text{h}dQ_\text{B}+V_\text{B}dP,
\end{align}
where the Legendre transformation is applied. Here, the change in the mass is clearly related to the pressure and volume. This cannot be seen from the first law without the $PV$ term. Therefore, according to an external field such as the charged scalar field, the change in the black hole should be balanced in consideration of the $PV$ term. Then, it will have physical implications that differ from the case without the $PV$ term.

\section{Solution to Charged Scalar Field Equation}\label{sec3}

Black holes can obtain conserved quantities, such as energy, momenta, and electric charge, by the scattering of an external field. The amount of conserved quantities taken into the black hole is given as the fluxes of the scattered external field. As much as the fluxes, the black hole can change its states while interacting with the external field. According to the fluxes, we can estimate the change in the black hole during an infinitesimal time interval. Here, we investigate the scattering of the nonminimally coupled massive scalar field with an electric charge to the charged AdS black hole in $D$-dimensional spacetime. Then, the solution to the charged scalar field is obtained at the outer horizon to obtain its fluxes. The action of the charged scalar field is
\begin{align}\label{eq:scalarlag01}
S_\Psi =-\frac{1}{2}\int d^D x \sqrt{-g}\left(\mathcal{D}_\mu \Psi \mathcal{D}^\mu \Psi^*+(\mu^2+\xi R)\Psi\Psi^*\right),
\end{align}
where the spacetime dimension is assumed to be $D\geq 4$. Owing to a scalar field with electric charge $q$, we consider the covariant derivative $\mathcal{D}_\mu=\partial_\mu-iq A_\mu$. The scalar field has the mass $\mu$ and nonminimal coupling $\xi$ with the curvature. There are two field equations, including the complex conjugate.
\begin{align}\label{eq:feqs01}
\frac{1}{\sqrt{-g}} \mathcal{D}_\mu \left(\sqrt{-g} g^{\mu\nu} \mathcal{D} _\nu \Psi\right)-(\mu^2+\xi R) \Psi=0,\quad \frac{1}{\sqrt{-g}} \mathcal{D}^*_\mu \left(\sqrt{-g} g^{\mu\nu} \mathcal{D}^* _{\nu} \Psi^*\right)-(\mu^2+\xi R) \Psi^*=0,
\end{align}
where we mainly focus on the solution to $\Psi$, because the solution to $\Phi^*$ is simply the complex conjugate to that of $\Psi$. According to Eqs.\,(\ref{eq:BHmetric01a}) and (\ref{eq:BHmetric01b}), the determinant of the metric is simply noted as
\begin{align}\label{eq:feqs02}
\sqrt{-g}=r^{D-2} \prod_{j=0}^{D-3}\sin^{D-2-j}\theta_j.
\end{align}
Then, substituting the gauge field in Eq.\,(\ref{eq:action02a}), the separable equation with respect to $\Psi$ is obtained as
\begin{align}\label{eq:feqs03}
\frac{1}{\sqrt{-g}}\partial_\mu(\sqrt{-g}g^{\mu\nu} \partial_\nu \Psi)-2iqA_0 g^{00} \partial_0 \Psi -q^2 g^{00}(A_0)^2 \Psi -(\mu^2+\xi R)  \Psi=0.
\end{align}
The solution to the scalar field is
\begin{align}\label{eq:solution05}
\Psi(t,r,\theta,\phi)=e^{-i\omega t}R(r) Y_{lm}(\theta_1,\theta_2,...\theta_{D-2}),
\end{align}
where $Y_{lm}(\theta_1,\theta_2,...\theta_{D-2})$ is the hyperspherical harmonics on a $(D-2)$-dimensional sphere\cite{Ida:2002zk}. The hyperspherical harmonics has its eigenvalue given as $-l(l+D-3)$. Then, the field equation in Eq.(\ref{eq:feqs03}) is separated into radial and angular parts. The radial equation is\cite{Konoplya:2003ii}
\begin{align}\label{eq:radialeq07}
-\frac{r^2}{\Delta}\left(-i\omega + iq \frac{Q}{r^{D-3}}\right)^2 R-\frac{l(l+D-3)}{r^2}R + \frac{(D-4)}{r^3}\Delta \partial_r R+\frac{\partial_r (\Delta \partial_r R)}{r^2}-(\mu^2+\xi \mathcal{R})  R=0,
\end{align}
where information regarding the angular momentum of the scalar field is simply compressed into the eigenvalue of the hyperspherical harmonics in Eq.\,(\ref{eq:radialeq07}). Since we consider a static black hole, the detailed values of the angular momentum of the scalar field are not important to our analysis. Note that the angular equations can be recurrently written in terms of the Laplace--Beltrami operator.
\begin{align}
\nabla^2_{\theta_i} Y_{ml}=(\sin\theta_i)^{2+i-D} \frac{\partial }{\partial \theta_i}\left((\sin\theta_i)^{D-2-i} \frac{\partial }{\partial \theta_i}Y_{ml}\right)+\frac{1}{\sin^2 \theta_i} \nabla^2_{\theta_{i+1}} Y_{ml},
\end{align}
where $\nabla^2_{\theta_i}$ is the Laplace--Beltrami operator on the $(D-1-i)$-dimensional sphere. 

The nontrivial behavior of the charged scalar field comes from the radial solution to Eq.\,(\ref{eq:radialeq07}). The radial solution can be obtained in simple form in the tortoise coordinate defined as
\begin{align}
\frac{dr^*}{dr}=\frac{r^2}{\Delta},
\end{align}
where the radial range of $r_\text{h}\leq r < +\infty$ becomes that of $-\infty< r^* \leq 0$. Then, the radial equation in Eq.\,(\ref{eq:radialeq07}) is rewritten in the tortoise coordinate as
\begin{align}\label{eq:radialeq09}
\frac{d^2 R}{d {r^*}^2}+\frac{(D-2)\Delta}{r^3}\frac{d R}{d {r^*}}+\left(\left(\omega-\frac{qQ}{r^{D-3}}\right)^2-\frac{\Delta}{r^2}\left((\mu^2+\xi R) r^2 +\frac{l(l+D-3)}{r^2}\right)\right)R=0.
\end{align}
Here, we need to consider the fluxes of the charged scalar field entering inside of the black hole through its outer horizon. Hence, the radial solution at the outer horizon provides the fluxes. In the limit of $r\rightarrow r_\text{h}$, the radial equation in Eq.\,(\ref{eq:radialeq09}) becomes a Schr\"{o}dinger-like equation.
\begin{align}
\frac{d^2 R}{d {r^*}^2}+\left(\omega-\frac{qQ}{r_\text{h}^{D-3}}\right)^2R=0,
\end{align}
where the mass and nonminimal coupling term of the scalar field in Eq.\,(\ref{eq:radialeq09}) is removed owing to $\Delta(r_\text{h})=0$. Then, the electric interaction only contributes to the radial solution. The radial solution of the scalar field at the outer horizon is
\begin{align}\label{eq:radialsol21}
R(r)= e^{\pm i \left(\omega - \frac{qQ}{r_\text{h}^{D-3}}\right)r^*}.
\end{align}
The ingoing field is the minus sign in Eq.\,(\ref{eq:radialsol21}), and the outgoing field is the plus sign. We select the former to represent the scalar field entering the outer horizon under the scattering. Therefore, the solutions of the two scalar fields are obtained as
\begin{align}\label{eq:scalarsolution01}
\Psi=e^{-i\omega t} e^{- i \left(\omega - \frac{qQ}{r^{D-3}_\text{h}}\right)r^*} Y_{lm}(\theta_1,\theta_2,...\theta_{D-2}),\quad \Psi^*=e^{i\omega t}  e^{i \left(\omega - \frac{qQ}{r^{D-3}_\text{h}}\right)r^*} Y^*_{lm}(\theta_1,\theta_2,...\theta_{D-2}).
\end{align}
Then, the exact forms of the scalar fields at the outer horizon can be estimated from Eq.\,(\ref{eq:scalarsolution01}). According to the solutions, transferred fluxes into the black hole can be obtained through the energy-momentum tensor of the scalar field.

\section{Thermodynamics under Charged Scalar Field}\label{sec4}

Here, we investigate changes in a charged AdS black hole owing to ingoing fluxes of the scattered scalar field. Flowing into the black hole, the energy and electric charge of the scalar field will vary as much as those of the black hole. Hence, depending on the energy and charge, properties of the black hole undergo changes that are expected to satisfy a specific relation between them. We show the relation between the conserved quantities of a black hole and the scalar field, considering the $PV$ term. The carried energy and electric charge of the scalar field are given as their fluxes at the outer horizon. These fluxes are obtained from the energy-momentum tensor that
\begin{align}
T^\mu_\nu=\frac{1}{2}\mathcal{D}^\mu\Psi \partial_\nu \Psi^*+\frac{1}{2}\mathcal{D}^{*\mu}\Psi^* \partial_\nu \Psi-\delta^\mu_\nu\left(\frac{1}{2}\mathcal{D}_\mu\Psi \mathcal{D}^{*\mu}\Psi^* -\frac{1}{2}(\mu^2+\xi R)\Psi\Psi^*\right).\nonumber
\end{align}
The energy flux is the component $T^r_t$ integrated by a solid angle on an $S^{D-2}$ sphere at the outer horizon. Further, we read the electric charge flux from the energy flux\cite{Bekenstein:1973mi}. Then, fluxes of energy and electric charge are
\begin{align}\label{eq:flux1a}
\frac{dE}{dt}=\int T_t^r \sqrt{-g}d\Omega_{D-2}=\omega\left(\omega-\frac{qQ}{r_\text{h}^{D-3}}\right)r_\text{h}^{D-2},\quad
\frac{de}{dt}=\frac{q}{\omega}\frac{dE}{dt}=q\left(\omega-\frac{qQ}{r_\text{h}^{D-3}}\right)r_\text{h}^{D-2}.
\end{align}
Fluxes in Eq.\,(\ref{eq:flux1a}) will infinitesimally change the corresponding properties of the black hole during the infinitesimal time interval $dt$. The electric charge flux obviously corresponds to the change in that of the black hole. The energy flux is ambiguous, however, because it can correspond to the enthalpy or internal energy of the black hole. Here, we will relate this to the {\it internal energy}. There are three reasons for doing so. First, when we consider thermodynamics without the $PV$ term, the internal energy consistently corresponds to the mass of the black hole, because $M_\text{B}=U_\text{B}$, compared with Eq.\,(\ref{eq:enthalpy01}). Second, this well reproduces the first law of thermodynamics with the $PV$ term, such that the choice ensures no loss of energy. Third, under this choice, the validity of the WCC conjecture is possible (regarding which validity, see the next section). By contrast, the alternative choice does not ensure these points. Hence, the changes in internal energy and electric charge are given as
\begin{align}\label{eq:fluxes17a}
dU_\text{B}=\left(\frac{dE}{dt}\right)dt,\quad dQ_\text{B}=\left(\frac{de}{dt}\right)dt.
\end{align}
Further, the change in the enthalpy is connected to the change in internal energy.
\begin{align}\label{eq:changeenthalpy01}
dU_\text{B}=d\left(M_\text{B}-PV_\text{B}\right)=\omega\left(\omega-\frac{qQ}{r_\text{h}^{D-3}}\right)r_\text{h}^{D-2}dt.
\end{align}
In consideration of the $PV$ term, ingoing fluxes of the external scalar field change both the mass and the volume of the black hole. Hence, this change is expected to differ considerably from the case without the $PV$ term. Note that the fluxes in Eq.\,(\ref{eq:flux1a}) are negative when
\begin{align}\label{eq:super10aq}
\omega < \frac{qQ}{r_\text{h}^{D-3}}.
\end{align}
Then, the black hole emits energy and charge through the charged scalar field under Eq.\,(\ref{eq:super10aq}). This is called superradiance, an interesting phenomenon observed in the scattering of an external field.  

Since we focus on thermodynamics and the WCC conjecture, the location of the outer horizon plays a significant role in our analysis. The outer horizon $r_\text{h}$ is located at the point satisfying $\Delta=0$, which has parameters $(M_\text{B},Q_\text{B},r_\text{h},\ell)$. When we assume that the parameters change to $(M_\text{B}+dM_\text{B},Q_\text{B}+dQ_\text{B},r_\text{h}+dr_\text{h},\ell+d\ell)$ by the scalar field, the changed location of the outer horizon is determined by 
\begin{align}\label{eq:delta01a}
\Delta(M_\text{B}+dM_\text{B},Q_\text{B}+dQ_\text{B},r_\text{h}+dr_\text{h},\ell+d\ell)=\frac{\partial \Delta_\text{h}}{\partial M_\text{B}}dM_\text{B}+\frac{\partial \Delta_\text{h}}{\partial Q_\text{B}}dQ_\text{B}+\frac{\partial \Delta_\text{h}}{\partial r_\text{h}}dr_\text{h}+\frac{\partial \Delta_\text{h}}{\partial \ell}d\ell=0,
\end{align}
where
\begin{align}
\Delta_\text{h}&\equiv \Delta{|}_{r=r_\text{h}}=0, \quad \frac{\partial \Delta_\text{h}}{\partial M_\text{B}}=-\frac{16\pi }{(D-2)r_\text{h}^{D-5}},\quad \frac{\partial \Delta_\text{h}}{\partial Q_\text{B}}=\frac{16\pi Q}{(D-2)r_\text{h}^{2D-8}},\quad \frac{\partial \Delta_\text{h}}{\partial \ell}=\frac{2r_\text{h}^4}{\ell^3},\\
\frac{\partial \Delta_\text{h}}{\partial r_\text{h}}&=2r_\text{h}-\frac{(2D-8)Q^2}{r_\text{h}^{2D-7}}+\frac{(2D-10)M}{r_\text{h}^{D-4}}+\frac{4r_\text{h}^3}{\ell^2}.\nonumber
\end{align}
One might notice that pressure is assumed to change in the scattering. We focus on pressure as a thermodynamic conjugate of the volume rather than the AdS radius. Then, the pressure can be varied such that it is balanced to the change in the volume of the black hole. Consequently, we consider the change in the pressure. However, we show that the change in the black hole is an {\it isobaric process}. This implies $d\ell=0$. Thus, {\it the pressure is constant}. In combination with Eqs.\,(\ref{eq:changeenthalpy01}) and (\ref{eq:delta01a}), the change in the outer horizon is obtained as
\begin{align}\label{eq:changeinouterhorizon01}
dr_\text{h}=\frac{16\pi r_\text{h}^{3}\ell^2 \left(\omega-\frac{qQ}{r_\text{h}^{D-3}}\right)^2}{\Omega_{D-2}(D-2)(d\Delta_\text{h} \ell^2-(D-1)r_\text{h}^3)}dt.
\end{align}
Here, even if we assume $d\ell=0$ in Eq.\,(\ref{eq:delta01a}), the change in the outer horizon is still derived by Eq.\,(\ref{eq:changeinouterhorizon01}). Hence, we can determine that the change in the black hole from the scattering of the charged scalar field is an isobaric process in terms of thermodynamics. Insofar as it is an isobaric process, {\it we set $d\ell=0$ without loss of generality in the following equations.} Moreover, the parameters $\omega$ and $q$ of the scalar field positively contribute to the change in the outer horizon of Eq.\,(\ref{eq:changeinouterhorizon01}). Then, we can show that the change is mainly determined by the initial state related to the denominator in Eq.\,(\ref{eq:changeinouterhorizon01}).

\begin{figure}[h]
\centering
\subfigure[{$\sigma$ graph in $D=4$.}] {\includegraphics[scale=0.61,keepaspectratio]{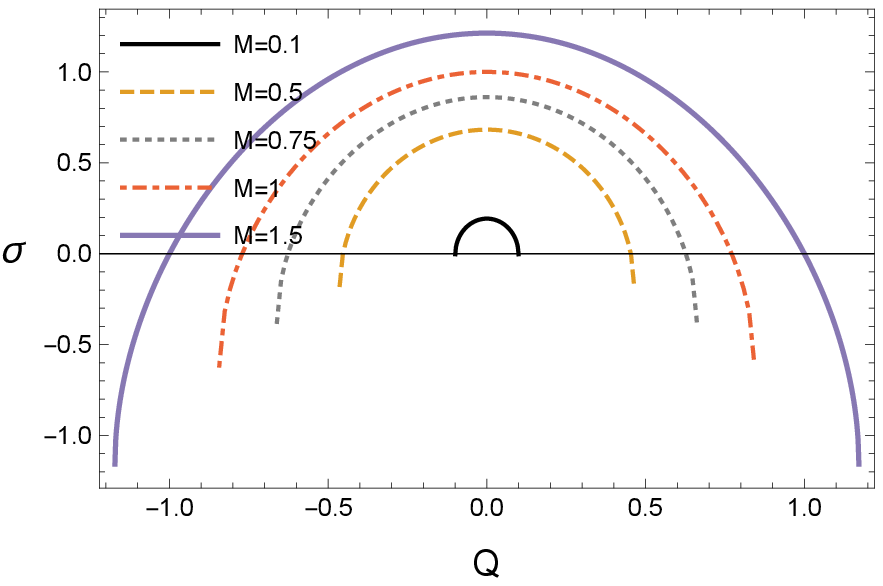}}\quad
\subfigure[{$\sigma$ graph in $D=5$.}] {\includegraphics[scale=0.6,keepaspectratio]{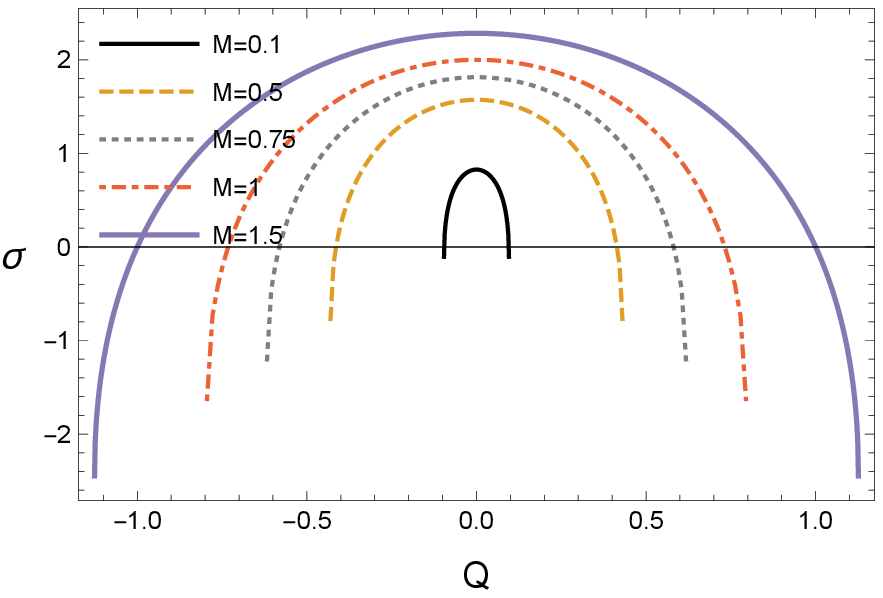}}\quad
\subfigure[{$\sigma$ graph in $D=6$.}] {\includegraphics[scale=0.6,keepaspectratio]{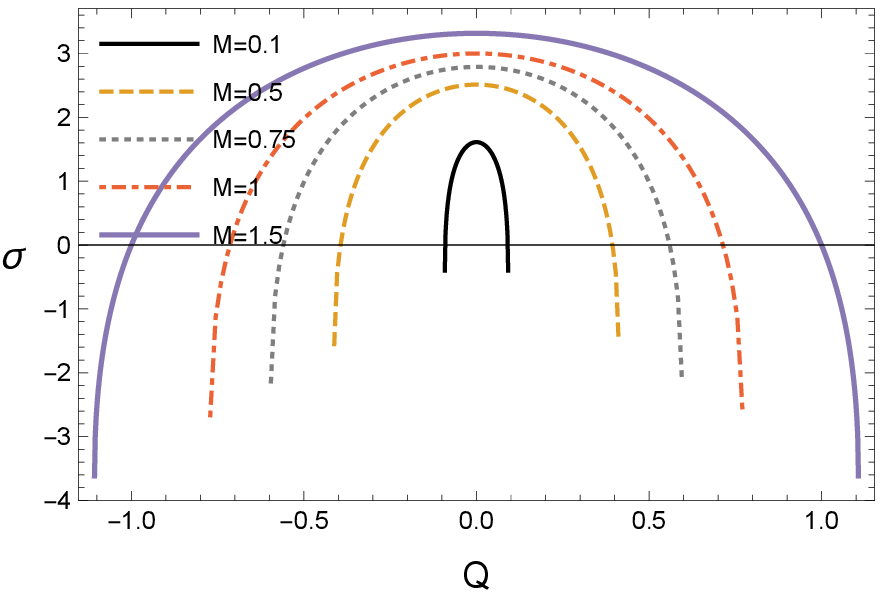}}
\caption{{\small Denominator $\sigma$ graphs regarding $Q$ with $\ell=1$ in various dimensions.}}
\label{fig:fig1sigma}
\end{figure}

\begin{figure}[h]
\centering
\subfigure[{$\sigma$ diagram in $D=4$.}] {\includegraphics[scale=0.5,keepaspectratio]{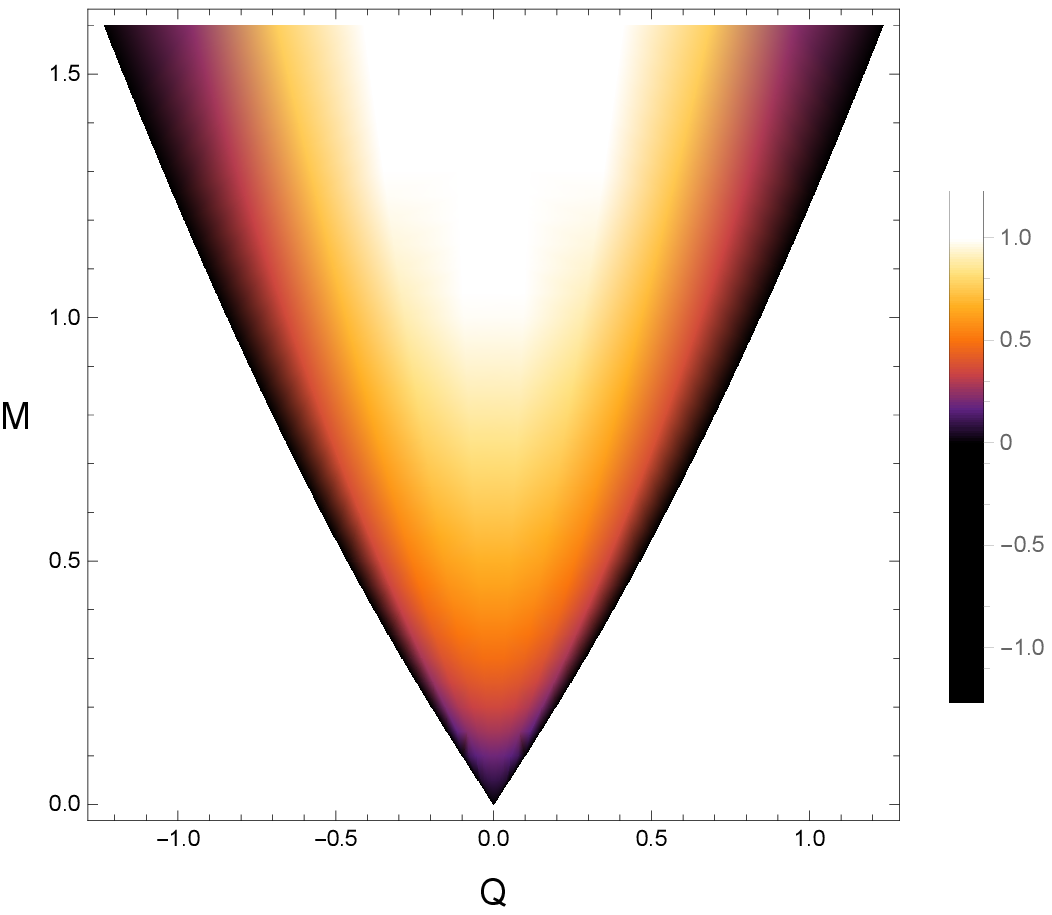}}\quad
\subfigure[{$\sigma$ diagram in $D=5$.}] {\includegraphics[scale=0.5,keepaspectratio]{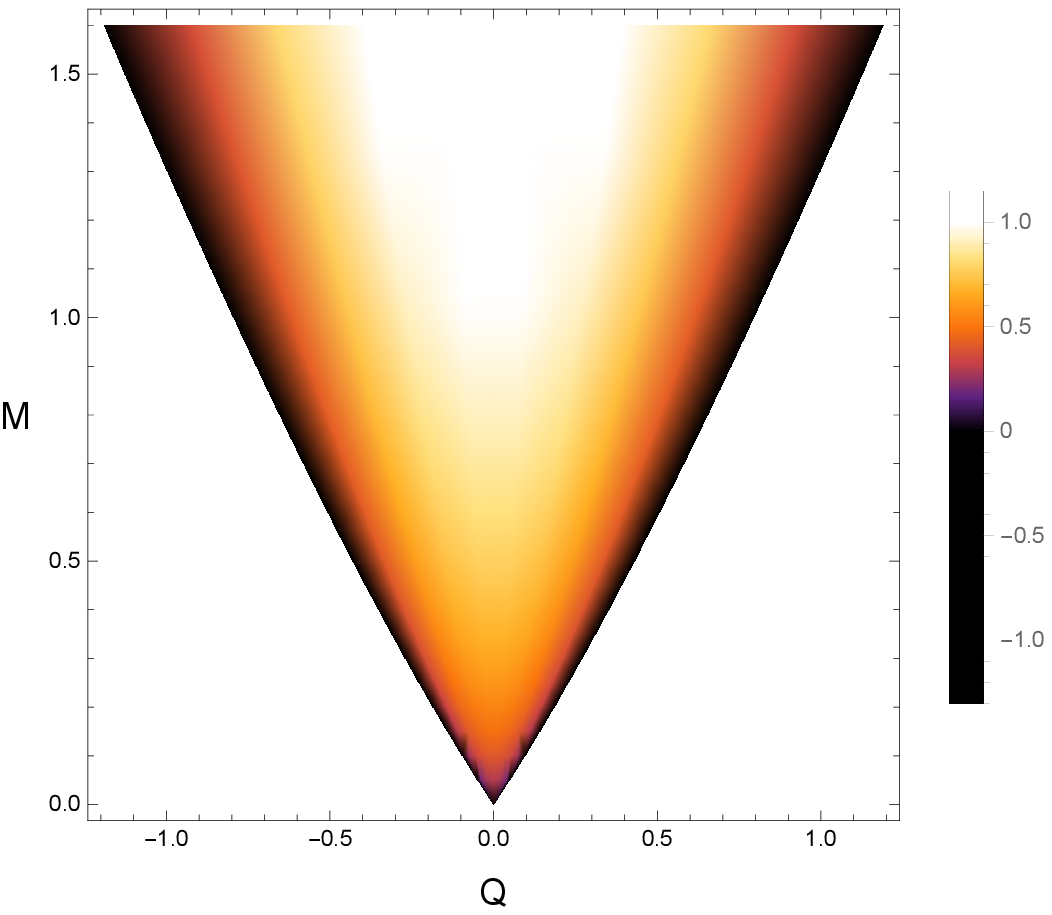}}\quad
\subfigure[{$\sigma$ diagram in $D=6$.}] {\includegraphics[scale=0.5,keepaspectratio]{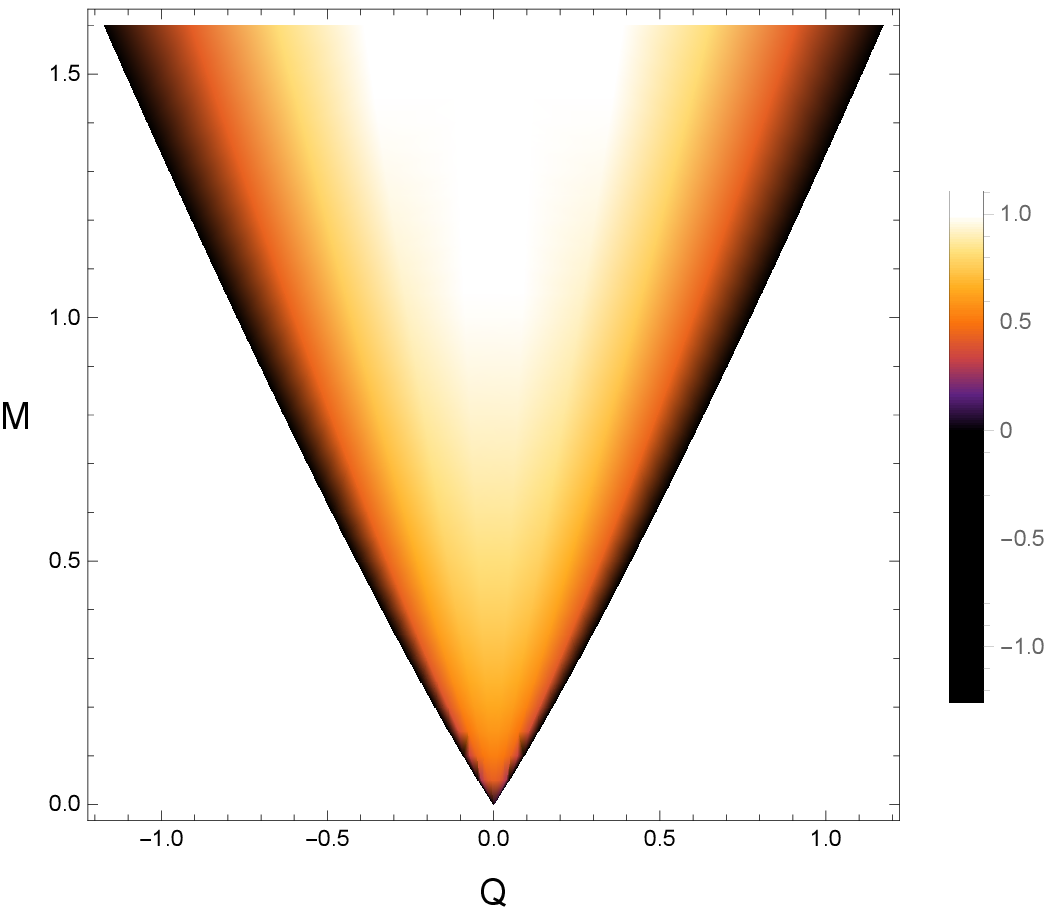}}
\caption{{\small Normalized denominator $\sigma$ in $Q$--$M$ diagrams with $\ell=1$ in various dimensions.}}
\label{fig:fig2phases}
\end{figure}

From the change in the outer horizon, we expect that Bekenstein--Hawking entropy behaves like Eq.\,(\ref{eq:changeinouterhorizon01}).
\begin{align}\label{eq:entropy02}
d S_\text{h}=\frac{1}{4}\Omega_{D-2}(D-2)r_\text{h}^{D-3}dr_\text{h},\quad dS_\text{h}=\frac{4\pi r_\text{h}^{D}\ell^2 \left(\omega-\frac{qQ}{r_\text{h}^{D-3}}\right)^2}{d\Delta_\text{h} \ell^2-(D-1)r_\text{h}^3}dt.
\end{align}
The change in entropy depends on the initial state, per Eq.\,(\ref{eq:changeinouterhorizon01}) with respect to the infinitesimal time $dt$. In particular, the denominator determines whether the entropy increases or decreases. Analytically, when the initial state is assumed for the extremal black hole, it satisfies $d\Delta_\text{h}=0$. Hence, the denominator can be rewritten as
\begin{align}\label{eq:sigmaextremal}
\sigma\equiv d\Delta_\text{h} \ell^2-(D-1)r_\text{h}^3=-(D-1)r_\text{h}^3<0,
\end{align}
which is always {\it negative} in any dimensions. This implies that the entropy of the extremal black hole can decrease owing to the scattering of the charged scalar field. This violates the second law of thermodynamics with the $PV$ term. Further, such a violation is firstly observed in consideration of the $PV$ term under the scattering of the charged scalar field. Details regarding the behavior of the denominator are shown numerically in Fig.\,\ref{fig:fig1sigma}. 
As we obtain Eq.\,(\ref{eq:sigmaextremal}), the denominator becomes negative with proximity to extremal black holes in Fig.\,\ref{fig:fig1sigma}. Further, this behavior occurs in any $D$-dimensional case. Hence, when the initial black hole is highly charged, the entropy decreases under the scattering of the scalar field. In other words, the initial state plays an important role in the change in entropy. Note that there are singular points of the change in the entropy, owing to $\sigma=0$ in near-extremal points in Fig.\,\ref{fig:fig1sigma}. Since the initial state $(M,Q)$ determines the signs of the change in the entropy, we can determine the range within which the second law of thermodynamics is violated, as shown in Fig.\,\ref{fig:fig2phases}.
Here, only the sign of the change is important, so we normalize the value of the denominator $\sigma$ between $-1$ and $1$. For a given mass, the denominator becomes negative when the initial state approaches a near-extremal one. Further, as the mass increases, the negative range widens. The range of the negative sign is narrow in higher-dimensional black holes, but it still exists in near-extremal and extremal black holes. Therefore, when we consider the $PV$ term, the violation of the second law can be observed in arbitrary dimensions.

However, the first law of thermodynamics is clearly observed in our analysis. In combination with Eqs.\,(\ref{eq:changeenthalpy01}) and (\ref{eq:entropy02}), we can obtain the first law with the infinitesimal change in the internal energy. In particular, the term for the outer horizon is divided into the change in entropy and $PV$ term with the Hawking temperature in Eq.\,(\ref{eq:temperature03a}). Then, the internal energy is obtained as
\begin{align}
dU_\text{B}=T_\text{h}dS_\text{h}+\Phi_\text{h}dQ_\text{B}-PdV_\text{B}.
\end{align}
According to the Legendre transformation, we can rewrite the internal energy into the enthalpy of the black hole. Hence, the first law of thermodynamics is ensured in terms of
\begin{align}
dM_\text{B}=T_\text{h}dS_\text{h}+\Phi_\text{h}dQ_\text{B}+V_\text{B}dP,
\end{align}
which implies that the thermodynamic energy of the system is conserved in consideration of the $PV$ term.

\section{Weak Cosmic Censorship in Near-Extremal and Extremal \\Charged AdS Black Holes}\label{sec5}

The WCC conjecture assumes that the singularity of a black hole should be hidden by the horizon from an asymptotic observer. This ensures that there is no naked singularity in the geometry of a black hole with a stable horizon. Here, we investigate the stability of the outer horizon under the scattering of the charged scalar field in consideration of thermodynamic pressure and volume. From the scattering and the isobaric process, the initial state $(M_\text{B}, Q_\text{B}, r_\text{h})$ becomes the final state $(M_\text{B}+dM_\text{B}, Q_\text{B}+dQ_\text{B}, r_\text{h}+dr_\text{h})$ during the infinitesimal time interval $dt$. Since the outer horizon is determined to be $\Delta(M_\text{B}, Q_\text{B}, r)=0$, we can estimate the existence of the horizon by testing solutions in $\Delta(M_\text{B}+dM_\text{B}, Q_\text{B}+dQ_\text{B}, r)=0$. This process can be simplified to an analysis of the change in the minimum value of $\Delta$.
 \begin{figure}[h]
\centering
\subfigure[{$\Delta$ in nonextremal black holes.}] {\includegraphics[scale=0.60,keepaspectratio]{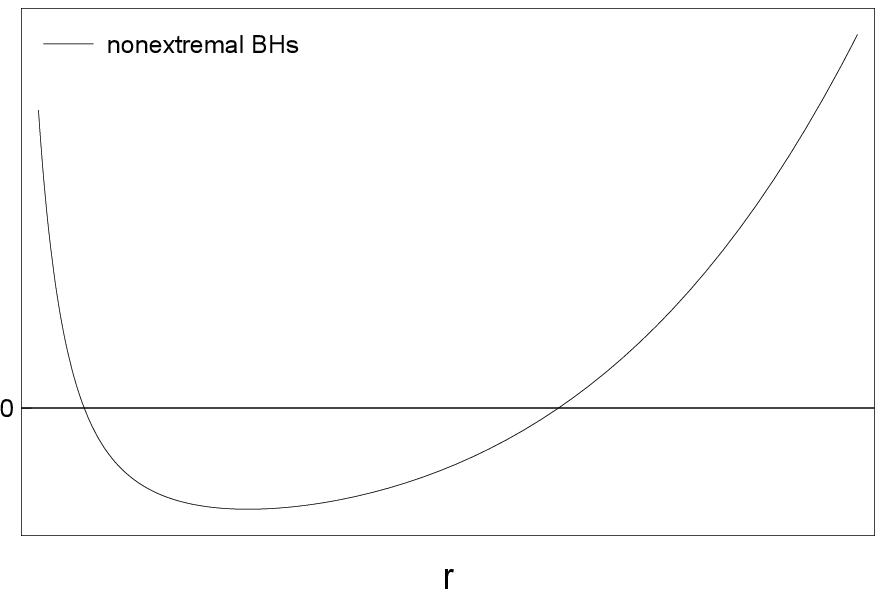}}\quad
\subfigure[{$\Delta$ in extremal black holes.}] {\includegraphics[scale=0.60,keepaspectratio]{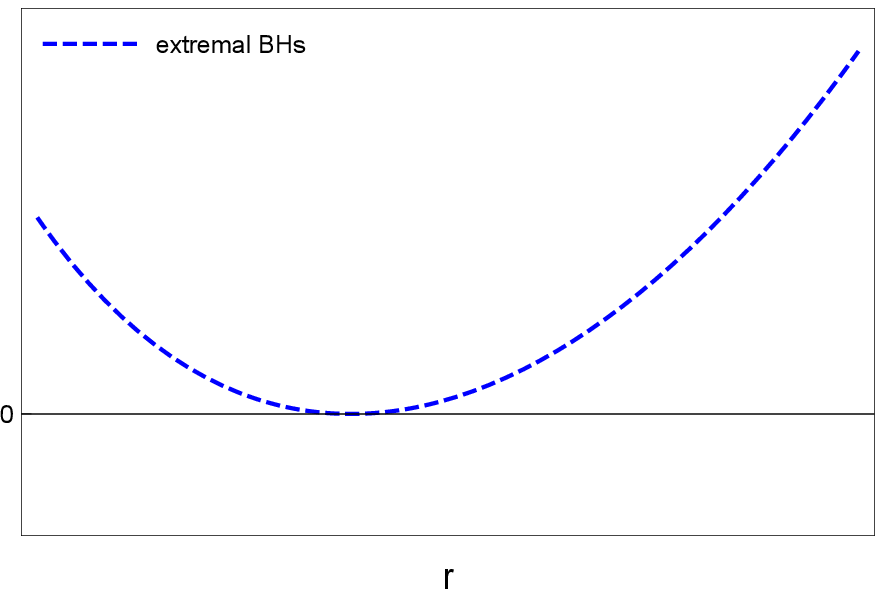}}\quad
\subfigure[{$\Delta$ in naked singularities.}] {\includegraphics[scale=0.60,keepaspectratio]{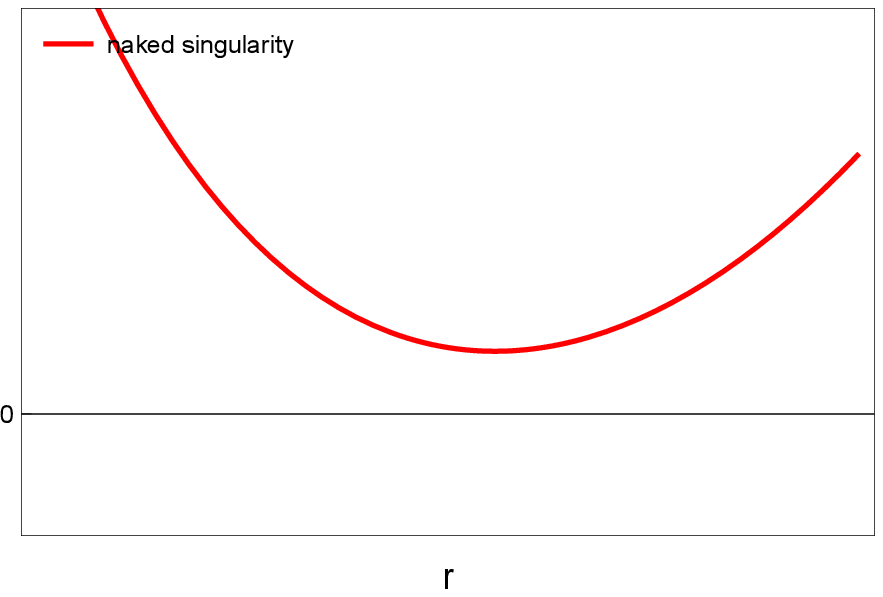}}
\caption{{\small Graphs of $\Delta$ for given states of the charged AdS black holes.}}
\label{fig:fig3phases}
\end{figure}
As shown in Fig.\,\ref{fig:fig3phases}, we begin with the initial state in a near-extremal or extremal black hole in Fig.\,\ref{fig:fig3phases}\,(a) or (b). In this initial state, the minimum value of the function $\Delta$ is negative or zero. As such, there exist solutions to their corresponding horizons. If the fluxes of the scalar field enter the black hole, the mass and electric charge of the black hole change owing to that of the charged scalar field during the infinitesimal time $dt$, and the minimum value also varies according to changes in the mass and charge of the black hole. Under the changes, if the minimum value is positive, as shown in Fig.\,\ref{fig:fig3phases}\,(c), no solution represents a horizon in the final state, and the black hole thus becomes a naked singularity. In this case, we can conclude that the black hole is overcharged, and that the WCC conjecture is invalid. Contrarily, in the final state, other cases, such as Fig.\,\ref{fig:fig3phases}\,(a) or (b), imply that the horizon still stably covers the inside of the black hole. Consequently, an asymptotic observer cannot see the singularity and the WCC conjecture is thus valid. Therefore, the sign of the minimum value in the final state is the key to validating the WCC conjecture in the scattering of the charged scalar field. 

Here, we investigate the sign of the minimum value of the function $\Delta$ in the final state. The sign can be obtained in terms of the initial state, because the final state is infinitesimally different from the initial state as much as the transferred conserved charges by the fluxes during the infinitesimal time interval $dt$. This infinitesimal change becomes significant when the initial state is a {\it near-extremal or extremal black hole}, rather than a nonextremal one. The minimum value of a near-extremal black hole ({\it including} an extremal one) is very close to zero. It thus has the possibility of being positive by infinitesimal changes contributing to the external scalar field. Therefore, we focus on a near-extremal black hole as an initial state. The near-extremal condition of the initial state is given at the minimum point $r_\text{min}$ with a negative constant $|\delta| \ll 1$ representing the minimum value of $\Delta$.
\begin{align}\label{eq:cosmicinitial2a}
\Delta_\text{min}&\equiv \Delta(r_\text{min})= r_\text{min}^2-\frac{2M}{r_\text{min}^{D-5}}+\frac{Q^2}{r_\text{min}^{2D-8}}+\frac{r_\text{min}^4}{\ell^2}=\delta\leq 0,\\
\frac{\partial \Delta_\text{min}}{\partial r_\text{min}}&=2r_\text{min}+\frac{2(D-5)M}{r_\text{min}^{D-4}}-\frac{2(D-4)Q^2}{r_\text{min}^{2D-7}}+\frac{4r_\text{min}^3}{\ell^2}=0,\nonumber\\
\frac{\partial ^2\Delta_\text{min}}{\partial r_\text{min}^2}&=2-\frac{2(D-4)(D-5)M}{r_\text{min}^{D-3}}+\frac{2(D-4)(2D-7)Q^2}{r_\text{min}^{2D-6}}+\frac{12r_\text{min}^2}{\ell^2}>0.\nonumber
\end{align}
Then, beginning with the initial state given in Eq.\,(\ref{eq:cosmicinitial2a}), we can estimate the infinitesimal change to the minimum value of the final state after an infinitesimal time interval $dt$. As a result of the fluxes, the mass and electric charge of the black hole change in the final state, and the minimum location $r_\text{min}$ moves to $r_\text{min}+dr_\text{min}$, which is obtained in terms of the initial state as
\begin{align}\label{eq:minimumvalue01a}
\Delta(M_\text{B}+dM_\text{B},Q_\text{B}+dQ_\text{B},r_\text{min}+dr_\text{min})=\Delta_\text{min}+\frac{\partial \Delta_\text{min}}{\partial M_\text{B}}dM_\text{B}+\frac{\partial \Delta_\text{min}}{\partial Q_\text{B}}dQ_\text{B}+\frac{\partial \Delta_\text{min}}{\partial r_\text{min}}dr_\text{min},
\end{align}
where
\begin{align}
\frac{\partial \Delta_\text{min}}{\partial M_\text{B}}=-\frac{16\pi}{\Omega_{D-2}(D-2)r_\text{min}^{D-5}}\equiv d\Delta_\text{M},\quad \frac{\partial \Delta_\text{min}}{\partial Q_\text{B}}=\frac{16\pi Q}{\Omega_{D-2}(D-2)r_\text{min}^{2D-8}}\equiv d\Delta_\text{Q}.\nonumber
\end{align}
In combination with Eqs.\,(\ref{eq:changeinouterhorizon01}), (\ref{eq:cosmicinitial2a}), and (\ref{eq:minimumvalue01a}), we obtain the minimum value.
\begin{align}\label{eq:deltamin02b}
\Delta(M_\text{B}+dM_\text{B},Q_\text{B}+dQ_\text{B},r_\text{min}+dr_\text{min})=\delta+\frac{r_\text{h}^{D-2}d\Delta_\text{h}d\Delta_\text{M} \ell^2}{d\Delta_\text{h} \ell^2-(D-1)r_\text{h}^3}\left(\omega -q\Phi_\text{h}\right)\left(\omega - q\Phi_\text{eff}\right)dt,
\end{align}
where the effective potential is obtained as
\begin{align}
\Phi_\text{eff}&\equiv \frac{Q ((D-1) r_\text{h}^6 r_\text{min}^D + r_\text{h}^D r_\text{min}^3 (d\Delta_\text{h} \ell^2-(D-1) r_\text{h}^3))}{d\Delta_\text{h} \ell r_\text{min}^D r_\text{h}^D}
.\nonumber
\end{align}
Since the initial state is a near-extremal black hole, the minimum location approaches the location of the outer horizon. Hence, we can draw a relation using the constant $\epsilon\ll 1$. Further, the constant $\delta$ can also be rewritten in terms of $\epsilon$ as
\begin{align}\label{eq:epsilondefine03a}
r_\text{h}\equiv r_\text{min}+\epsilon,\quad\delta=-\epsilon d\Delta_\text{h}+\mathcal{O}(\epsilon^3),
\end{align}
where we will show $d\Delta_\text{h}\sim \epsilon$. When the initial state is an extremal black hole, the minimum point is coincident with the location of the horizon, so $\epsilon=0$. Further, the minimum value of the initial state begins at zero. This is consistently denoted by $\delta=0$, as shown in Eq.\,(\ref{eq:epsilondefine03a}). In terms of $\epsilon$, the effective potential is 
\begin{align}\label{eq:solomegaeff03a}
\Phi_\text{eff}&=\Phi_\text{h}+\frac{Q(D-3)(d\Delta_\text{h} \ell^2-(D-1)r_\text{h}^3)}{r_\text{h}^{D-2}d\Delta_\text{h} \ell^2}\epsilon + \mathcal{O}(\epsilon^2),\\
d\Delta_\text{h}&=\left(2-\frac{2(D-4)(D-5)M}{r_\text{h}^{D-3}}+\frac{2(D-4)(2D-7)Q^2}{r_\text{h}^{2D-6}}+\frac{12r_\text{h}^2}{\ell^2}\right)\epsilon+ \mathcal{O}(\epsilon^2).\nonumber
\end{align}
For the validity of the WCC conjecture, the sign of the minimum value is significant in the final state. The horizon exists for a negative minimum value  in Eq.\,(\ref{eq:minimumvalue01a}). The right-hand side of Eq.\,(\ref{eq:minimumvalue01a}) can be rewritten as
\begin{align}\label{eq:inequaleq01a}
\Delta(M_\text{B}+dM_\text{B},Q_\text{B}+dQ_\text{B},r_\text{min}+dr_\text{min})=\delta-\frac{d\Delta_\text{M}r_\text{h}^{D-2}d\Delta_\text{h} \ell^2}{(D-1)r_\text{h}^3}\left(\omega -q\Phi_\text{h}\right)\left(\omega - q\Phi_\text{eff}\right)dt,
\end{align}
where we consider $d\Delta_\text{M}<0$ and $d\Delta_\text{h} \ell^2 \ll (D-1) r_\text{h}^3$. If the initial state is an extremal black hole, we can set $\delta,\epsilon=0$. Moreover, $d\Delta=0$ in the extremal case. Then, we can obtain the change to the minimum value from Eq.\,(\ref{eq:inequaleq01a}).
\begin{align}\label{eq:extremaltest1a}
\Delta(M_\text{B}+dM_\text{B},Q_\text{B}+dQ_\text{B},r_\text{min}+dr_\text{min})=0.
\end{align}
This implies that an extremal black hole is still an extremal black hole. The mass and electric charge of the black hole can change owing to the fluxes of the scalar field, but its state still satisfies the extremal condition. Therefore, the WCC conjecture is {\it valid} for extremal black holes. If the initial state is a near-extremal black hole, the scale of the variables becomes important. In our analysis, the time interval is assumed to be infinitesimally small. Hence, we can properly assume that the time interval is the same scale as $\epsilon$, so $dt\sim \epsilon$. Then, the change to the minimum value is obtained as
\begin{align}\label{eq:nearextremaltest1a}
\Delta(M_\text{B}+dM_\text{B},Q_\text{B}+dQ_\text{B},r_\text{min}+dr_\text{min})-\Delta(M_\text{B},Q_\text{B},r_\text{min})=\mathcal{O}(\epsilon^2),
\end{align}
where we consider Eq.\,(\ref{eq:solomegaeff03a}) and $q^2\ll 1$. Thus, the change to the minimum value can be assumed to be zero in the first order of $\epsilon$. This implies that there is no change to the state of the black hole: a near-extremal black hole is still near-extremal despite having different mass and electric charge. Therefore, the WCC conjecture is also {\it valid} for near-extremal black holes. Note that when the second order of $\epsilon$ is considered in Eq.\,(\ref{eq:nearextremaltest1a}), a near-extremal black hole reaches a slightly extremal state in $\omega/q >\Phi_\text{h}$. Likewise, it reaches a slightly nonextremal state in $\omega/q <\Phi_\text{h}$. However, owing to Eq.\,(\ref{eq:extremaltest1a}), the black hole cannot be overcharged under the scattering of the charged scalar field. Thus, the WCC conjecture is {\it valid}.

When testing the WCC conjecture, we must recognize the importance of the energy and charge transferring through the fluxes of the scalar field during the time interval $dt$. When the time interval is introduced in our analysis, we have to assume that the energy and charge of the scalar field flow into the black hole in infinitesimally small pieces during $dt$. This clearly differs from particles entering black holes in previous studies \cite{Toth:2011ab,Gao:2012ca,Gwak:2016gwj,Chirco:2010rq}, where the WCC conjecture is invalid in near-extremal black holes, because the particle can transfer conserved quantities to overcharge the black hole beyond the extremal condition. To resolve this issue in the particle, we need to consider that the conserved quantities of the particle are absorbed into the black hole in infinitesimally small pieces\cite{Gwak:2016gwj}. However, with the scattering of the scalar field, infinitesimally small energy and charge are transferred into the black hole during this infinitesimal time $dt$. Thus, the concept of absorbing infinitesimally small pieces is already inherent in the time interval. Hence, there is no overcharging in the scattering. This is an important feature for the validity of the WCC conjecture with the charged scalar field.

\section{Superradiance with Pressure and Volume}\label{sec6}

When we consider the $PV$ term in a charged AdS black hole, we already show that the changes in the black hole differ from those without the $PV$ term. These changes suggest the possibility of a different evolution of the black hole in the time flow from each time interval $dt$, depending on whether or not the $PV$ term is considered. According to Eq.\,(\ref{eq:flux1a}), the fluxes of the scattered scalar field depend on the electric potential of the black hole. When $\omega/q > Q/r^{D-3}$, the fluxes are positive, so the energy and charge flow into the black hole. Moreover, when $\omega/q < Q/r^{D-3}$, the negative fluxes represent energy and charge flowing out of the black hole. This superradiance is an interesting phenomenon observed in the scattering of the scalar field. According to absorption and superradiance, the state of the black hole can be expected to be saturated to an equilibrium where $\omega/q = Q/r^{D-3}$. Here, we assume that the scalar field has a frequency that is infinitesimally smaller than the electric potential of the black hole. Further, the ratio $\omega/q$ of the scalar field remains constant. Then, 
\begin{align}\label{eq:kappa01a}
\frac{\omega}{q}=\Phi_\text{h}-\epsilon,
\end{align}
where $\epsilon\ll 1$. Hence, the fluxes in Eq.\,(\ref{eq:flux1a}) become negative, inducing superradiance. Since the fluxes are very small, however, changes to the black hole occur slowly. Thus, we investigate whether the black hole can be saturated into the equilibrium in a finite period of time. During a time interval $dt$, the initial electric potential infinitesimally changes 
\begin{align}\label{eq:deltaphi03a}
\Phi_\text{h}(Q_\text{B}+dQ_\text{B},r_\text{h}+dr_\text{h})&=\Phi_\text{h}+\frac{\partial \Phi_\text{h}}{\partial Q_\text{B}}dQ_\text{B}+\frac{\partial \Phi_\text{h}}{\partial r_\text{h}}dr_\text{h}=\Phi_\text{h}+d\Phi_\text{h},
\end{align}
where
\begin{align}
\frac{\partial \Phi_\text{h}}{\partial Q_\text{B}}=\frac{8\pi }{(D-2)\Omega r_\text{h}^{D-3}},\quad \frac{\partial \Phi_\text{h}}{\partial r_\text{h}}=-\frac{(D-3)Q}{r^{D-2}}.\nonumber
\end{align}
Substituting Eqs.\,(\ref{eq:changeenthalpy01}), (\ref{eq:changeinouterhorizon01}), and (\ref{eq:kappa01a}) into Eq.\,(\ref{eq:deltaphi03a}), the change in the electric potential $d\Phi_\text{h}$ is obtained in terms of the first-order $\epsilon$ as
\begin{align}\label{eq:domega17}
d\Phi_\text{h}=-\frac{8\pi q^2 r_\text{h}dt}{(D-2)\Omega_{D-2}}\epsilon+\mathcal{O}(\epsilon^2).
\end{align}
Then, since $d\Phi_\text{h}/\epsilon\ll 1$, even if $\epsilon$ is infinitesimally small, sufficient and considerable time is needed to saturate the potential of the black hole to $\omega/q$ of the scalar field. In particular, the scale of $q$ is much smaller than that of the black hole: $q\ll Q$. Thus, as the time-step $dt$ proceeds, the electric potential of the black hole approaches the ratio $\omega/q$, though saturation cannot be achieved during one time-step. Instead, it requires a very long time to obtain. Moreover, when the black hole absorbs energy and charge, we can observe the same behavior. Under transformation $\epsilon \rightarrow -\epsilon$ in Eq.\,(\ref{eq:kappa01a}), we obtain 
\begin{align}\label{eq:domega18}
d\Phi_\text{h}=\frac{8\pi q^2 r_\text{h}dt}{(D-2)\Omega_{D-2}}\epsilon+\mathcal{O}(\kappa^2).
\end{align}
This further implies that saturation owing to absorption requires considerable time, as denoted in Eq.\,(\ref{eq:domega17}). Therefore, in both cases, we conclude that a charged AdS black hole slowly saturates its electric potential to the external scalar field.

\section{Summary}\label{sec7}

We investigated the laws of thermodynamics and the WCC conjecture under the scattering of a nonminimally coupled massive scalar field with an electric charge in a $D$-dimensional charged AdS black hole. We considered thermodynamic pressure and volume, and expected distinct behavior depending on whether the $PV$ term was considered. Since changes in a black hole occur according to the transferred energy and electric charge from the scalar field, the amount of energy and charge is estimated from the fluxes of the solution to the $D$-dimensional scalar field at the outer horizon where the contributions of mass and nonminimal coupling in the scalar field are removed, leaving only the charge effect. The infinitesimal changes in the black hole should be well-related to each other in terms of the first law of thermodynamics when regarding enthalpy as the mass of a black hole. Indeed, the first law is observed when we assume that the energy flux of the scalar field contributes to the {\it internal energy} of a black hole in consideration of its consistency with the WCC conjecture. However, the second law of thermodynamics is violated slightly in the near-extremal and extremal range. This is only observed with the $PV$ term. Under scattering, changes to the mass and electric charge of the black hole induce changes in the function $\Delta$, which determines the existence of the horizons. For the WCC conjecture, we estimated the change in the function $\Delta$ with respect to an infinitesimal time interval. According to the change to the minimum value of $\Delta$, extremal black holes remain extremal despite a different mass and charge contributing to the fluxes. Likewise, near-extremal black holes remain near-extremal in the first order because the fluxes of the scalar field have a limit to the transferred energy and charge to the black hole during the time interval. Thus, the WCC conjecture is {\it valid}. Here, regarding the validity of the WCC conjecture, the limits to transferred energy and charge owing to the time interval play an important role in preventing the transfer of a large amount of energy and charge to the black hole and thus preventing it from being overcharged beyond the extremal condition. We also tested whether a black hole can reach an equilibrium with a fixed scalar field in consideration of the $PV$ term. Our results indicate that as the time-step proceeds, the electric potential of a black hole approaches equilibrium. However, the saturation of the potential to the scalar field requires a very long time, because the time derivative of the potential becomes small. Therefore, when we consider the $PV$ term, the WCC conjecture is valid, but changes in extremal and near-extremal black holes are distinct in their detailed responses with respect to the scattering of the charged scalar field.

\vspace{10pt} 

\noindent{\bf Acknowledgments}

\noindent This work was supported by the National Research Foundation of Korea (NRF) grant funded by the Korea government (MSIT) (NRF-2018R1C1B6004349). B.G. appreciates APCTP for its hospitality during completion of this work.

\end{document}